\pdfoutput=1

\documentclass{sig-alternate-2013}

\usepackage{xspace}
\usepackage{numprint}
\usepackage{subcaption}
\usepackage{algorithm,algorithmic}
\usepackage{booktabs}
\usepackage{mathtools}
\usepackage{url}
\usepackage{hyperref}
\usepackage{xspace}
\usepackage{color}
\usepackage{amssymb}
\usepackage{latexsym}
\usepackage{multirow}

\newcommand{\scterm}[1]{\textsc{#1}\xspace}
\newcommand{\txt}[1]{\textsf{#1}\xspace}

\newcommand{\imdb}{IMDb\xspace}

\begin{document}
%
% --- Author Metadata here ---
%\CopyrightYear{2007} % Allows default copyright year (20XX) to be over-ridden - IF NEED BE.
%\crdata{0-12345-67-8/90/01}  % Allows default copyright data (0-89791-88-6/97/05) to be over-ridden - IF NEED BE.
% --- End of Author Metadata ---

\title{``Roles for the boys?'' Mining cast lists for gender and role distributions over time}
\numberofauthors{2}
\author{
% 1st. author
\alignauthor Will Radford\\
       \affaddr{Xerox Research Centre Europe}\\
       \affaddr{6 chemin de Maupertuis}\\
       \affaddr{38240 Meylan, France}\\
       \email{will.radford@xrce.xerox.com}
% 2nd. author
\alignauthor Matthias Gall\'e\\
       \affaddr{Xerox Research Centre Europe}\\
       \affaddr{6 chemin de Maupertuis}\\
       \affaddr{38240 Meylan, France}\\
       \email{matthias.galle@xrce.xerox.com}
}

\maketitle
\begin{abstract}
Film and television play an important role in popular culture, however studies that require watching and annotating video are time-consuming and expensive to run at scale.
We explore information mined from media database cast lists to explore onscreen gender depictions and how they change over time.
We find differences between web-mediated onscreen gender proportions and those from US Census data.
We propose these methodologies are a useful adjunct to traditional analysis that allow researchers to explore the relationship between online and onscreen gender depictions.
\end{abstract}

%A category including the fourth, optional field follows...



\section{Introduction}
Film and television are an integral part of culture and one way that people understand and interact with it.
Onscreen scenarios reflect the values from some real or imagined story, but also inform the viewers expectations.
However, attempting to directly study film and television presents some issues.
Watching video for analysis does not scale well to large datasets without significant manual effort.
This limits most large-scale study to easily digestible data sources: film popularity, box-office figures, reviews, scripts and other metadata.
Although non-video data sources may be easier to study, they limit the types of questions researchers can ask.
For example, box office figures do not allow detailed analysis of cinematography.

\begin{figure}
  \centering
  \fbox{\includegraphics[width=.42\textwidth]{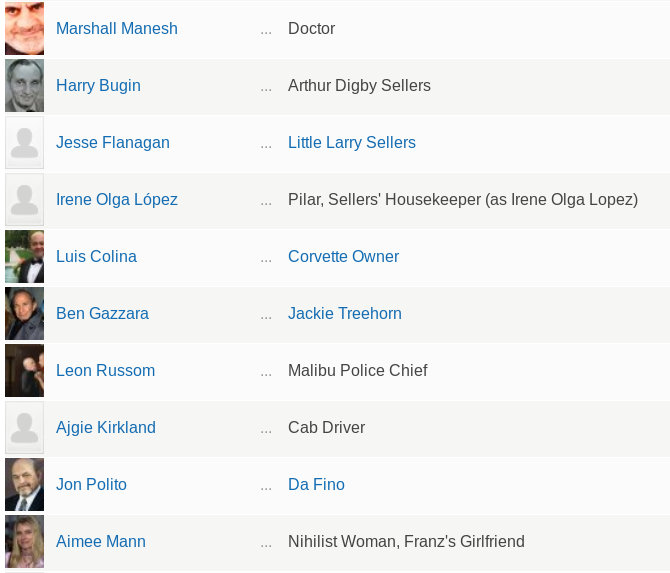}}
  \caption{Excerpt from the cast list for ``The Big Lebowski''.}
  \label{cast-list}
\end{figure}

Our research question is whether web science can provide viable proxies that let us answer interesting social science research questions at scale.
We use data available from a popular film and television website and examine \emph{cast lists}.
Figure \ref{cast-list} is a section of the Internet Movie Database (\imdb)\footnote{Alexa ranking 49 (global), 24 (US) as of 22/1/15.} cast list from ``The Big Lebowski''\footnote{\url{www.imdb.com/title/tt0118715}}, showing performer names and images on the left, with their character name on the right.
Some character names are names (e.g. \txt{Arthur Digby Sellers}), but some are professional roles (e.g. \txt{Doctor}) or combinations of role and relation to other characters (e.g. \txt{Nihilist Woman, Franz's Girlfriend}).
We exploit three factors from the data: productions are listed with their release date, male and female performers are distinguished in the data, and unnamed characters are usually listed by their role or profession.
This lets us count gendered performances of a particular role over time, which can be used to explore social science questions.

This paper is structured as follows: we discuss related work in media gender studies and \imdb in Section \ref{s:rw}.
Section \ref{s:data} describes the dataset and the methodology we use to handle noisy user-generated data.\footnote{Code at \url{https://github.com/wejradford/castminer}}
We then explore what roles are found onscreen and how that changes over time in Section \ref{s:roles}.
In Section \ref{s:gender}, we examine how roles interact with gender over time and how this compares to real-world gender distributions.
We believe that web science methodologies can augment traditional manual analysis to enable comparison of online and onscreen gender depictions.

\newpage
\section{Background}
\label{s:rw}

Gender is a complex sociocultural phenomenon with a vast academic literature and we stress that this work makes limited exploration of gender itself.
Instead we focus on some of the issues relating to gender in media as much as our data allows.
Under-representation of women is a long-standing gender issue in media, both in terms of the gender of performers and also the subject matter, for example proportions of news stories that focus on females \cite{Wood94}.
Moreover, Wood notes stereoptypical portrayals of hypermasculine, yet domestically incompetent, male characters and the female characters dependent on them, and complex relationships of power and image.
This trend is confirmed in a more recent meta-study of articles in a special issue of the \emph{Sex Roles} journal \cite{Collins11}, which adds to this observations about the role of race and interesting conjecture about the effect of under-representation and the importance of also finding positive representations of women in media.

Many of gender media research questions require manual analysis.
In their study of screen portrayals and media employment, Smith et al. consider \numprint{26225} characters\footnote{\numprint{4506} of these were speaking roles.} from the 600 top-grossing films from 2007--2013 \cite{Smith14}.
They find a low percentage of female speaking characters -- consistently around 30\% over each year of their sample, and only 2\% of films features more female than male characters.
They also study sexualisation of female characters, finding them more likely to be shown in revealing clothing, nude or referred to as attractive.
They note the dearth of female content creators, noting that the number of female writers and directors is at a six year low circa 2014.
This extensive and detailed study is only made possible with a team of 71 highly-trained student coders and to apply this depth of research at scale would be difficult and costly.

\imdb is an interesting source of data due to its size and popularity on the internet.
Boyle notes that ``\imdb has been the focus of surprisingly little academic attention'' in her study of gender and movie reviews \cite{Boyle14}.
This consisted of analysing how gender is expressed (or not) in textual reviews for three different films and the online profiles of the reviewers.
Data from \imdb has been used for research in the natural language processing and computational linguistics domain, primarily as the source of a corpus of movie reviews annotated with sentiment \cite{pang-lee-vaithyanathan:2002:EMNLP02}.
Other resources for gender information have been gathered from the US Census and automatically processed web text \cite{Bergsma:05,bergsma-lin:2006:COLACL}.
A possible application for gender data is in coreference resolution \cite{pradhan-EtAl:2011:CoNLL-ST}, the task of clustering \emph{mentions} that refer to the same entity in a document.
For example, lists of male and female names may provide evidence whether the mentions \txt{he}, \txt{Bob} and \txt{manager} should be matched together.

%\begin{figure}[t!]
%  \centering
%  \includegraphics[width=0.49\textwidth]{role_counts}
%  \caption{Count of roles over time.}
%  \label{role_counts}
%\end{figure}

Detailed gender analyses of media are compelling yet difficult to conduct at scale.
We hope to use metadata about screen media as a proxy for the original media to explore, albeit in a limited way, issues about gender and its onscreen representation.
Web science methodologies, such as those used to study scanned books \cite{Michel14012011}, suggest useful starting points.
The dataset in this study allows us to study how people report onscreen media using the web, but this kind of data can also influence other media.
Specifically, cast information is part of the ecosystem of media reporting, advertising, review and commentary, and this can have real-world impact.
A study focussing on the dynamics of online film reviews found that volume significantly impacts box office sales, rather than content and ratings \cite{journals/dss/DuanGW08}.
The authors attribute this to an indicator of underlying word-of-mouth information flow and that online reviews spread awareness of the film.
User data is increasingly being directly used to assist decisions about what media a studio should produce\footnote{\url{http://www.newyorker.com/business/currency/hollywoods-big-data-big-deal}} and this is indicative of the complex relationship between onscreen media and the web.

\section{Dataset and methods}
\label{s:data}

Our methodology requires two simplifying assumptions.
We assume that \imdb is a good proxy for onscreen entertainment, which we believe is a reasonable assumption for recent productions, but less so for older productions as we discuss below.
We also assume that popular film and television is more likely to appear in a database like \imdb, and as such its aggregated content is a good estimator of what a random person would watch.
Following from this, we ask the question: \emph{``What are viewers likely to learn about roles and gender over time from onscreen entertainment?''}.

We downloaded the plain text data files \texttt{actors.list.gz} and \texttt{actresses.list.gz}\footnote{Accessed on 24/10/14 from \url{http://www.imdb.com/interfaces}.} and applied several cleaning phases.
The files list the performer name and the titles and dates of productions they appear in.
Unfortunately, these lists do not distinguish between films, television, so it is difficult to distinguish between media -- clearly an important methodological question.
We exclude records where the performer is listed using an alternative name \txt{(as \ldots)}, and generate one record per appearance in a film or television episode.
We further process records based on the role, filtering roles marked \txt{n/a}, or those that reference selves (e.g. \txt{himself, herself or themselves}).
We also remove markers of multiple similar roles: ordinal prefixes (e.g. \txt{first} or \txt{1st}) from 1 to 5 and suffixes (e.g. \txt{(1)} or \txt{(\#1)}).
Finally, we remove any text in parentheses and split multi-role characters (e.g. \txt{model/actress}), generating one count for each lower-cased role.
We aggregate roles by year and calculate a gender distribution for each role $r$ and year $y$.
Specifically, $p(\text{F}|r,y)$ is the count of records with role $r$ in year $y$ by a performer from the actresses list, normalised by the count of all $r$ and $y$ records.\footnote{$p(\text{M}|r,y) =  1 - p(\text{F}|r,y)$.}

\begin{figure}[t!]
  \centering
  \includegraphics[width=0.49\textwidth]{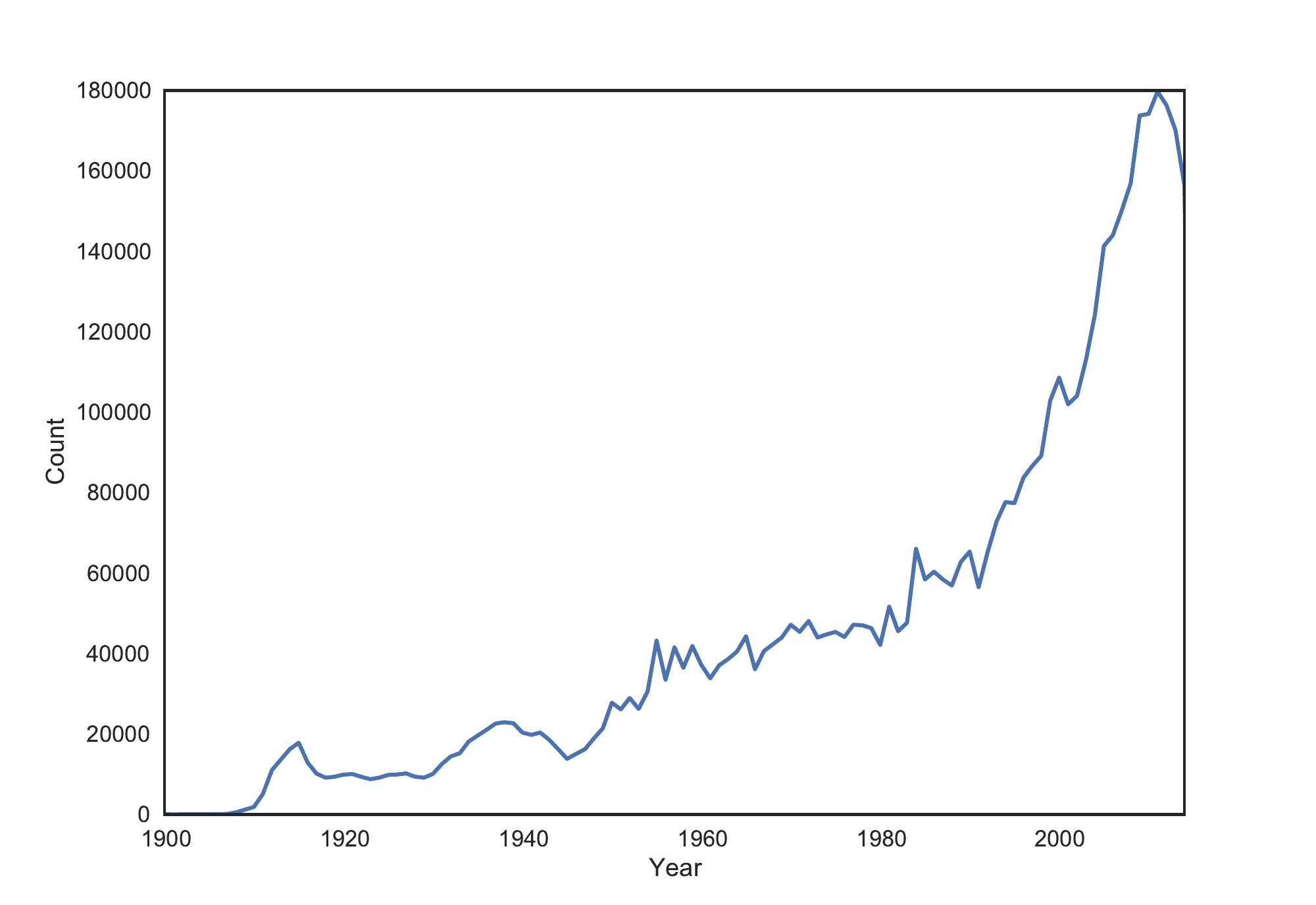}
  \caption{Count of roles over time.}
  \label{role_counts}
\end{figure}

As with most user-generated content, there are a number of caveats that apply to the data and our analysis.
It is possible that performers can be misclassified and added to the wrong list file, or records listed with incorrect years.
We would expect this to be the result of data entry error and focus our analysis on those with higher count, as to avoid this hopefully rare occurrence.
There is also a significant observation bias as while it may be common for film and television to be listed as it enters production today, older productions are only listed if a user takes the effort to document them.
As a result, older counts are susceptible to skew towards television productions with a strong internet-based community dedicated to listing each and every episode.

% This table needs to be here for the formatting to behave.
\begin{table*}
  \centering
  {\small
  \begin{tabular}{cccccc}
  \toprule
  1900-1920	&	1920-1940	&	1940-1960	&	1960-1980	&	1980-2000	&	2000-2020\\
  \midrule
  undetermined role	&	minor role	&	newsreader	&	host	&	host	&	host\\
  mary	&	henchman	&	host	&	model	&	hostess	&	contestant\\
  jack	&	reporter	&	reporter	&	announcer	&	newsreader	&	narrator\\
  the girl	&	dancer	&	narrator	&	presenter	&	presenter	&	presenter\\
  the wife	&	policeman	&	panelist	&	various	&	announcer	&	guest\\
  the sheriff	&	undetermined role	&	townsman	&	narrator	&	narrator	&	judge\\
  minor role	&	townsman	&	announcer	&	singer	&	guest	&	panelist\\
  the husband	&	detective	&	sports newsreader	&	guest	&	various	&	various characters\\
  policeman	&	party guest	&	singer	&	reporter	&	additional voices	&	hostess\\
  daughter	&	waiter	&	weather forecaster	&	various characters	&	reporter	&	reporter\\
  \bottomrule
  \end{tabular}
  }
  \caption{Top 10 roles for 20 year periods from 1920.}
  \label{role_double_decade_counts}
\end{table*}

\begin{table*}
  \centering
  {\small
  \begin{tabular}{cccccc}
  \toprule
  1900-1920	&	1920-1940	&	1940-1960	&	1960-1980	&	1980-2000	&	2000-2020\\
  \midrule
  undetermined role	&	henchman	&	newsreader	&	model	&	additional voices	&	zombie\\
  mary	&	reporter	&	host	&	various	&	anchor	&	housemate\\
  jack	&	dancer	&	panelist	&	various characters	&	contestant	&	police officer\\
  the girl	&	townsman	&	announcer	&	member of the short circus	&	musical director	&	alex\\
  the wife	&	waiter	&	sports newsreader	&	paul williams	&	lexicographer	&	laura\\
  the sheriff	&	narrator	&	weather forecaster	&	victor newman	&	interviewer	&	audience member\\
  minor role	&	barfly	&	corresponsal	&	brady black	&	ridge forrester	&	david\\
  the husband	&	doctor	&	correspondent	&	jack abbott	&	phil	&	bar patron\\
  policeman	&	bit role	&	presenter	&	george	&	emcee	&	sam\\
  daughter	&	bartender	&	sports reporter	&	roman brady	&	co-hostess	&	sarah\\
  \bottomrule
  \end{tabular}
  }
  \caption{Top 10 \textbf{newly popular} roles for 20 year periods from 1920.}
  \label{role_double_decade_counts_emerging}
\end{table*}

%Gender is a more sophisticated social construct than its binary modelling in this dataset, but in this case, we are constrained by the data available.
We do not distinguish between films and television, and our processing considers a television episode equal to a film.
This skews the data in favour of television and future work may be able to map to other resources to tease them apart.
Likewise, we do not distinguish between the production country, which rules out potentially interesting national comparisons and language processing.
We do not further process roles and so some may be character names and others professions.
We might expect that professions will have higher counts, as it is more likely that generic roles are repeated in many records than character names.
This means that we are comparing names and roles, which is somewhat inelegant, but extracting roles for main characters would require linking to external structured (e.g. Freebase) or unstructured plot data (e.g. Wikipedia).
Moreover, central characters are more important, but it's not immediately clear how to weight their influence so we believe that our approach is a pragmatic compromise.
If we were able to map to media country, the language-dependent processing would be possible.
This might include mapping \txt{host} and \txt{hostess} using stemming, but this comes at the cost of conflating dissimilar concepts within or across languages.
Finally, the role descriptions do not follow a fixed schema, so some equivalent role counts may be split by virtue of general synonymy (e.g. \txt{director} and \txt{filmmaker}) or different gender forms (e.g. \txt{policeman}, \txt{policewoman}, \txt{cop}, \txt{police officer}).
This problem may be alleviated by mapping \imdb roles onto a semantic ontology such as WordNet \cite{Miller95wordnet:a}.

After preprocessing, we retain \numprint{15468002} role records from between 1900 and 2020 (Figure~\ref{role_double_decade_counts_emerging}).
%and Figure \ref{role_counts} shows the count of records over time.
The number of entries grows from the early 20th century and increase steadily until the 1990s, when the rate of growth increases.
Note that, although the data was collected in 2014, there are records dated later than that, as \imdb lists ongoing and planned productions.\footnote{We consider all data for counts, but graphs do not show data after 2014.}

\begin{figure}[t!]
  \centering
  \includegraphics[width=0.49\textwidth]{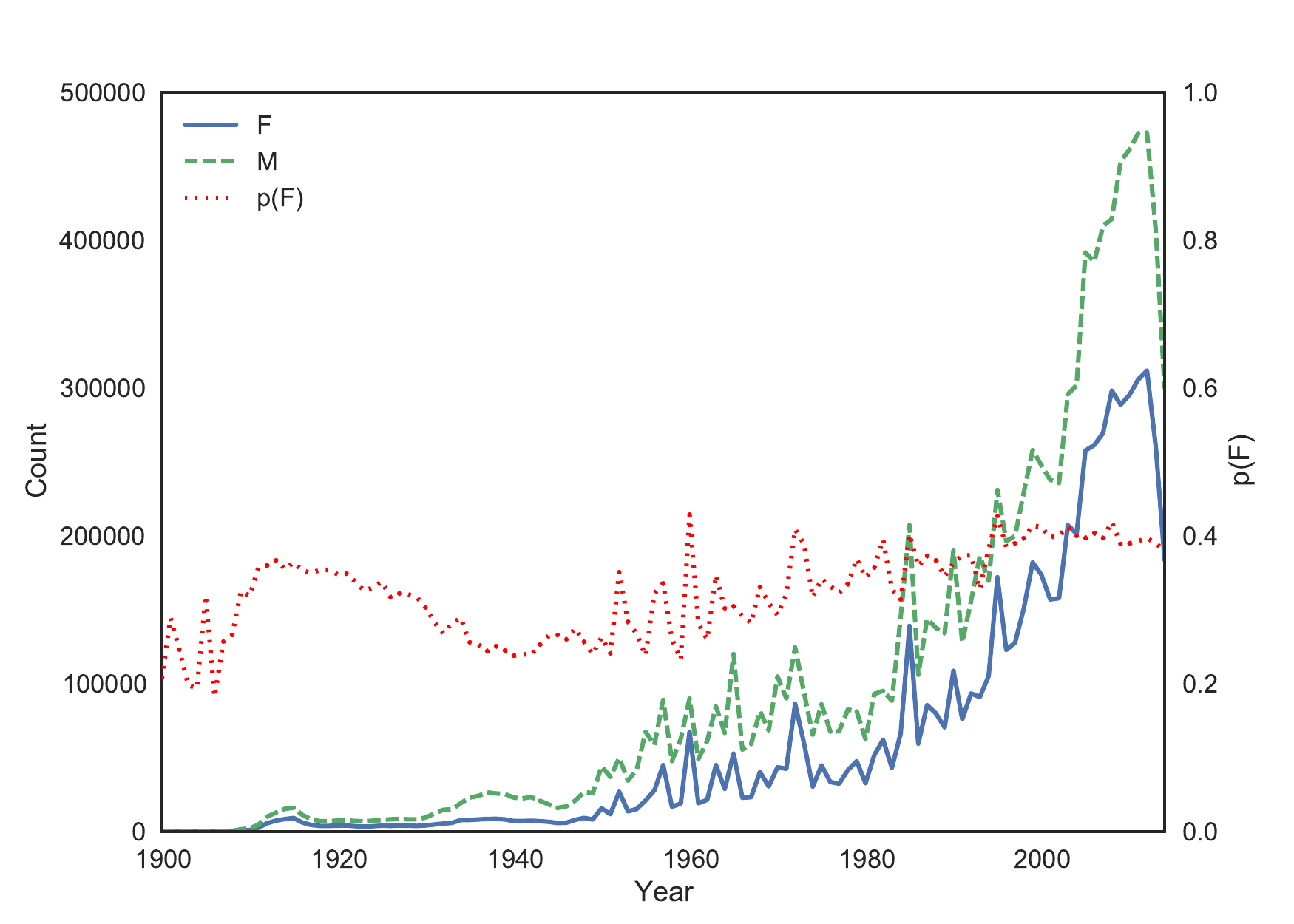}
  \caption{Count of roles from each gender over time, as well as the gender distribution $p(\text{F})$.}
  \label{gender_counts}
\end{figure}

\section{Roles}
\label{s:roles}
The dataset allows us to track, at a very coarse level, what roles are popular in onscreen media and how has this changed over time.
Table \ref{role_double_decade_counts} shows the top 10 most common roles in 20 year periods from 1900.
This shows how roles have changed over time and reflects what roles are reported and seen on screen.
Initial roles from 1900 are most often \txt{undetermined} or stock characters (\txt{mary}, \txt{jack}, \txt{the girl}, \txt{the wife}, \txt{daughter}, \txt{husband}).
Roles from 1920-1940 are made up of dramatic roles that appear to be drawn from a crime or noir genre: \txt{henchman}, \txt{policeman}, \txt{detective}.
Others are ambiguous, as \txt{reporter} and \txt{dancer} could either be in a dramatic or actual role in a news broadcast or variety show.
For the two decades from 1940, there seems to be a shift towards news broadcasting (i.e. \txt{newsreader}, \txt{sports newsreader}, \txt{weather forecaster}), narration (i.e. \txt{announcer}, \txt{narrator}) and hosted television with \txt{host}, \txt{singer} and \txt{panelist}.
The trend of hosted television is maintained for the rest of the dataset, but we see evidence of shifts in trend: \txt{model} from 1960--1980, \txt{additional voices} for cartoons from 1980--2000, and finally reality television roles from 2000 (i.e. \txt{contestant}, \txt{judge}).

While the above analysis shows the enduring popularity of hosted screen entertainment, this can obscure some of the emerging roles through time.
Table \ref{role_double_decade_counts_emerging} shows, for the same period, which roles are new and did not appear in the top 50 roles of the previous period.
The 1900s list is the same as Table \ref{role_double_decade_counts} as this is the first period used.
The 1920s sees different descriptions of underspecified roles (\txt{bit role} vs \txt{undetermined role}).
There is a strong focus on hosted and news media from the 1940s and evidence of non-English-speaking entries (\txt{corresponsal} is Spanish for \txt{correspondent}).
From the 1960s, there is evidence of popular roles in children's television (\txt{member of the short circus} from ``The Electric Company''), television soap operas (\txt{paul williams}, \txt{victor newman}\footnote{This character seems to first appear in 1980, so may be listed under an incorrect year. In lieu of canonical sources for ``The Young and the Restless'': \url{http://en.wikipedia.org/wiki/Victor_Newman}} from ``The Young and the Restless'').
Newly popular roles in the 1980s and 1990s included game and quiz shows (\txt{contestant}, \txt{lexicographer} from ``Countdown Masters''), different television soap operas (\txt{ridge forrester} from ``The Bold and the Beautiful'') and new terms (\txt{anchor} and the gendered form \txt{co-hostess}).
Roles thusfar from the two decades from 2000 reflects the recent trend for \txt{zombie} characters in television, driven in part by the success of productions such as ``The Walking Dead'', which typically feature many unnammed zombie characters and thus has a large impact on the count data.
We see a continued trend of more first-name roles (\txt{laura}, \txt{david} and the gender-ambiguous \txt{alex} and \txt{sam}), and roles that reflect current naming conventions (\txt{police officer} rather than \txt{policeman}, the generic role \txt{mother} and \txt{bar patron} rather than the earlier \txt{bar fly}).
One concern with this method is that by only considering roles that have not been seen in a previous top 50, then we may find that the listed roles are low rank or count with respect to the overall roles (i.e. as per Table \ref{role_double_decade_counts}).
The lowest rank was 40 (\txt{sarah} in 2000--2020) and the lowest count was 614 (\txt{bartender} in 1920--1940).

We propose that the dataset is an interesting way to explore how onscreen roles change over time.
We see evidence for a main hosted model of onscreen entertainment, with secondary trends, such as reality television.
In older performances there seems also to be evidence of a skew towards television programmes that have been comprehensively documented, presumably by a dedicated internet-based community.

\section{Gender}
\label{s:gender}

One of the most valuable characteristics of our dataset is that each performer has gender information.
Aggregating by role allows us to consider biases of the gender of onscreen roles.
Figure \ref{gender_counts} shows how roles over time are split between two genders, with counts for each gender and also the proportion of female roles ($p(F)$).
From 1940, we see a gradual increase in the proportion of roles played by female actors from 0.25 to 0.4.
Before this period, total counts are somewhat lower, so it is difficult to draw conclusions.
The higher female proportion around 1920 may reflect the fact that records correspond to film, not television, but this is difficult to establish without taking extra metadata into account.

\begin{table}[t!]
  \centering
  {\tiny
  \begin{tabular}{lrlr}
  \toprule
  Role	& F &	Role	& M \\
  \midrule
  host	& \numprint{123775}	& 	host	& \numprint{370187}	\\
  hostess	& \numprint{74856}	& 	narrator	& \numprint{75736}	\\
  presenter	& \numprint{39551}	& 	announcer	& \numprint{58356}	\\
  newsreader	& \numprint{34145}	& 	presenter	& \numprint{51762}	\\
  model	& \numprint{30289}	& 	guest	& \numprint{46107}	\\
  guest	& \numprint{29296}	& 	various	& \numprint{33917}	\\
  contestant	& \numprint{28651}	& 	newsreader	& \numprint{32289}	\\
  reporter	& \numprint{25911}	& 	various characters	& \numprint{31785}	\\
  nurse	& \numprint{20852}	& 	contestant	& \numprint{31739}	\\
  dancer	& \numprint{19039}	& 	reporter	& \numprint{31190}	\\
  panelist	& \numprint{17820}	& 	panelist	& \numprint{25999}	\\
  various	& \numprint{14541}	& 	judge	& \numprint{25036}	\\
  judge	& \numprint{14123}	& 	additional voices	& \numprint{22906}	\\
  narrator	& \numprint{13714}	& 	co-host	& \numprint{22177}	\\
  co-host	& \numprint{12314}	& 	doctor	& \numprint{18299}	\\
  various characters	& \numprint{12047}	& 	policeman	& \numprint{16590}	\\
  girl	& \numprint{11595}	& 	performer	& \numprint{15964}	\\
  singer	& \numprint{11509}	& 	man	& \numprint{13680}	\\
  woman	& \numprint{11197}	& 	bartender	& \numprint{13327}	\\
  waitress	& \numprint{11147}	& 	various roles	& \numprint{12522}	\\
  correspondent	& \numprint{10691}	& 	singer	& \numprint{12463}	\\
  mother	& \numprint{10009}	& 	correspondent	& \numprint{12356}	\\
  laura	& \numprint{9930}	& 	dancer	& \numprint{12173}	\\
  maria	& \numprint{9860}	& 	musical guest	& \numprint{11937}	\\
  additional	& \numprint{9652}	& 	waiter	& \numprint{11876}	\\
  performer	& \numprint{8582}	& 	police officer	& \numprint{11206}	\\
  sarah	& \numprint{8235}	& 	cop	& \numprint{10812}	\\
  lisa	& \numprint{8122}	& 	soldier	& \numprint{10185}	\\
  anna	& \numprint{8002}	& 	david	& \numprint{10087}	\\
  co-hostess	& \numprint{7847}	& 	student	& \numprint{10070}	\\
  student	& \numprint{7624}	& 	guard	& \numprint{9906}	\\
  mary	& \numprint{6960}	& 	detective	& \numprint{9720}	\\
  rita	& \numprint{6908}	& 	paul	& \numprint{9315}	\\
  alice	& \numprint{6744}	& 	tom	& \numprint{9124}	\\
  rosa	& \numprint{6730}	& 	sports newsreader	& \numprint{9078}	\\
  jane	& \numprint{6022}	& 	john	& \numprint{9068}	\\
  various roles	& \numprint{5922}	& 	jack	& \numprint{8978}	\\
  julie	& \numprint{5790}	& 	commentator	& \numprint{8864}	\\
  secretary	& \numprint{5692}	& 	mike	& \numprint{8536}	\\
  sara	& \numprint{5546}	& 	townsman	& \numprint{8522}	\\
  linda	& \numprint{5427}	& 	max	& \numprint{8508}	\\
  receptionist	& \numprint{5419}	& 	extra	& \numprint{8363}	\\
  extra	& \numprint{5221}	& 	frank	& \numprint{8281}	\\
  eva	& \numprint{5135}	& 	boy	& \numprint{8271}	\\
  marta	& \numprint{5013}	& 	mark	& \numprint{7999}	\\
  jenny	& \numprint{5002}	& 	tony	& \numprint{7936}	\\
  sandra	& \numprint{4930}	& 	george	& \numprint{7895}	\\
  ana	& \numprint{4860}	& 	musician	& \numprint{7840}	\\
  teresa	& \numprint{4800}	& 	interviewee	& \numprint{7822}	\\
  clara	& \numprint{4775}	& 	joe	& \numprint{7803}	\\
  \bottomrule
  \end{tabular}}
  \caption{The 50 most frequent female and male roles.}
  \label{tb:top_gender_roles}
\end{table}

\begin{table*}[t!]
  \centering
  {\tiny
  \begin{tabular}{lrlrlrlrlr}																				
  \toprule																				
  \multicolumn{2}{c}{Strongly male}	&			\multicolumn{2}{c}{Moderately male}	&	\multicolumn{2}{c}{Gender neutral}	&	\multicolumn{2}{c}{Moderately female}	&	\multicolumn{2}{c}{Stongly female} \\
  \midrule																				
  Role & $p(F)$&				Role & $p(F)$&				Role & $p(F)$&				Role & $p(F)$&				Role & $p(F)$\\
  \midrule																				
  delivery man  & 	0.00  &		band  & 	0.05  &		emt  & 	0.17  &		corresponsal  & 	0.35  &		member of the short circus  & 	0.60  \\
  color commentator  & 	0.00  &		little boy  & 	0.05  &		player  & 	0.18  &		center square  & 	0.35  &		secretary  & 	0.88	\\
  father  & 	0.00  &		basketball player  & 	0.07  &		additional voice  & 0.20  &		patient  & 	0.35  &		mother  & 	0.93  	\\
  boyfriend  & 	0.00  &		biker  & 	0.07  &		trainer  & 	0.22  &		co-host  & 	0.36  &		nurse  & 	0.94  \\
  policeman  & 	0.00  &		moderator  & 	0.09  &		host  & 	0.25  &		hotel guest  & 	0.36  &		old woman  & 	0.96 	\\
  musical director  & 	0.00  &		coroner  & 	0.10  &		mentor  & 	0.26  &		office worker  & 	0.40  &		model  & 	0.97 	\\
  truck driver  & 	0.01  &		fbi agent  & 	0.10  &		guest co-host  & 	0.27  &		news anchor  & 	0.42  &		actress  & 	0.98  	\\
  inspector  & 	0.02  &		bailiff  & 	0.11  &		inmate  & 	0.28  &		android  & 	0.43  &		maid  & 	0.98 	\\
  monk  & 	0.02  &		bartender  & 	0.13  &		passerby  & 	0.29  &		candidate  & 	0.44  &		stewardess  & 	0.99  	\\
  soldier  & 	0.02  &		staff humorist  & 	0.14  &		journalist  & 	0.31  &		participant  & 	0.47  &		secretaria  & 	1.00 	\\
  \bottomrule
  \end{tabular}																																		
  }
  \caption{Examples of common roles with different gender distributions.}
  \label{role_examples}
\end{table*}

Table~\ref{tb:top_gender_roles} shows the 50 most frequent roles per gender.
Of course, some of the roles of Table~\ref{role_double_decade_counts} appear again here, but it is already possible to see biases towards one of the genders.
\txt{model} and \txt{receptionist} are frequent roles which are mostly female, as are \txt{hostess}, \txt{girl}, \txt{woman}, \txt{waitress} and \txt{mother}, together with a series of frequent female first names.
On the male side side, there seems to be strong bias for \txt{narrator}, \txt{announcer}, \txt{doctor}, \txt{detective}, \txt{bartender} together with a series of security or military roles (\txt{police officer}, \txt{cop}, \txt{soldier}, \txt{guard}), and again some gender-specific roles like  \txt{policemen}, \txt{men}, \txt{boy}, \txt{waiter}.

We can also analyse the gender distribution of common roles to characterise how gender relates to roles at a high level.
As an example, we filtered the most common mentions with an overall count above \numprint{1000} that did not belong to a list of common names from the US Census.
To try and characterise the space of roles, we ordered them by $p(\text{F})$ and partitioned them into five equal bins and randomly sampled 10 entries from each.
Table \ref{role_examples} shows the results: on both extremes there are again gendered roles (\txt{boyfriend}, \txt{actress}), while more towards the middle section some more interesting biases can be observed (\txt{biker} and \txt{basketball player} as male and \txt{secretary} as female). 
Note that due to the overall higher count of male occurrences, the midpoint of gender distribution is between the ``moderately'' and ``strongly'' female classes.

\begin{table}[t!]
  \centering
  {\small
  \begin{tabular}{lcr}
  \toprule
  Profession 		&  	Keywords											& $p(F)$ \\
  %\hline
  %\hline
  \midrule
  IT			&	software, computer, hacker 							& 0.51 \\
  %\hline
  \multirow{2}{*}{Doctor}		&	medical, dr, dr., doctor & \multirow{2}{*}{0.23} \\
            & md, physician	& \\
  %\hline
  Corporate	&	corporate, ceo, coo									& 0.18 \\
  %\hline
  Law			&	prosecutor, lawyer									 & 0.15 \\
  %\hline
  \multirow{2}{*}{Politics}	&	minister, dictator, parlament & \multirow{2}{*}{0.09}  \\
            &  senator, president & \\
  %\hline
  Science		&	science, professor									& 0.09 \\
  %\hline
  \multirow{4}{*}{Religion}	&	priest, priestess, reverend & \multirow{4}{*}{0.08} \\
                            & pastor, prior, allamah  & \\
                            & imam, rabbi, guru, lama &  \\
                            & bishop, ayatollah, swami & \\

  %\hline
  Engineering & engineer												& 0.05 \\
  \bottomrule
  \end{tabular}
  }
  \caption{Gender distribution grouped by profession.}
  \label{tb:genderBias}
\end{table}

In~\cite{Smith14}, % gender bias report
the authors analyze 120 movies and show strong biases in the representation of executive roles. 
Inspired by that report, we looked for key roles in areas such as law, IT and religion and looked at the aggregated count of male and female actor in these roles.
For each keyword listed in Table~\ref{tb:genderBias}, we looked for all roles that contained that word.
We made exceptions for \txt{president} where we looked only for exact matches, and \txt{bishop} where we ignored those mentions that end with it to avoid including surnames.

Law and corporate professions had around 15\% of female representation, which coincides with the values reported in~\cite{Smith14} for Law but not for corporate professions, while the medical domain (doctors) had a female probability of $0.23$.
In contrast to the results in \cite{Smith14}, Religion does not score at the bottom with regards to female presentation (although very low with $0.08$). 
From the professions we selected, Engineering was the lowest (0.05).
The highest scoring profession was IT (0.52), which is partly due to the fact that many computer voices were female (\txt{computer} had 460 female occurrences, versus 247 male ones; and \txt{enterprise computer} from ``Star Trek'' was almost exclusively female).

\begin{figure*}[t!]
  \begin{subfigure}{\textwidth}
    \centering
    \includegraphics[width=\textwidth]{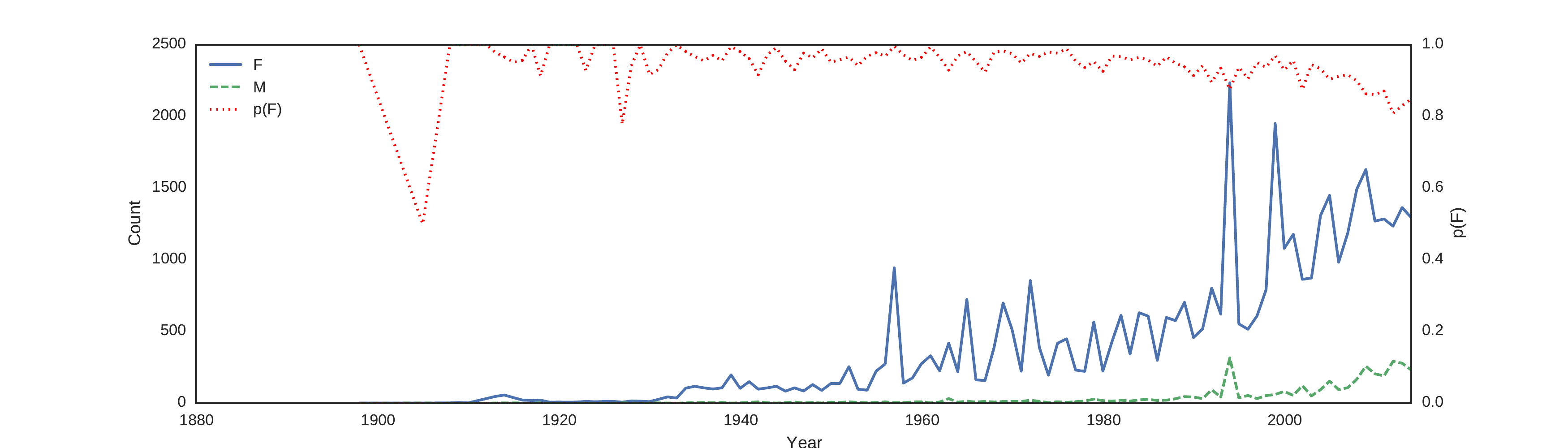}
    \caption{\txt{nurse}}
    \label{nurse}
  \end{subfigure}%

  \begin{subfigure}{\textwidth}
    \centering
    \includegraphics[width=\textwidth]{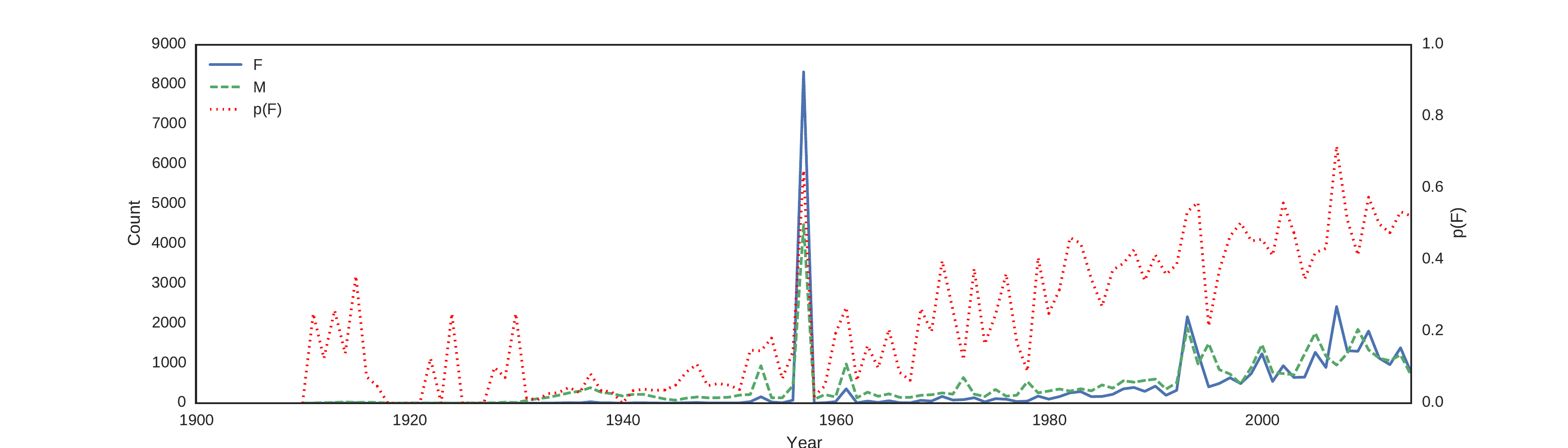}
    \caption{\txt{reporter}}
    \label{reporter}
  \end{subfigure}

  \caption{Gender counts and proportions over time for various roles.}
  \label{gender-roles}
\end{figure*}

\begin{figure}[t!]
  \centering
  \includegraphics[width=0.45\textwidth]{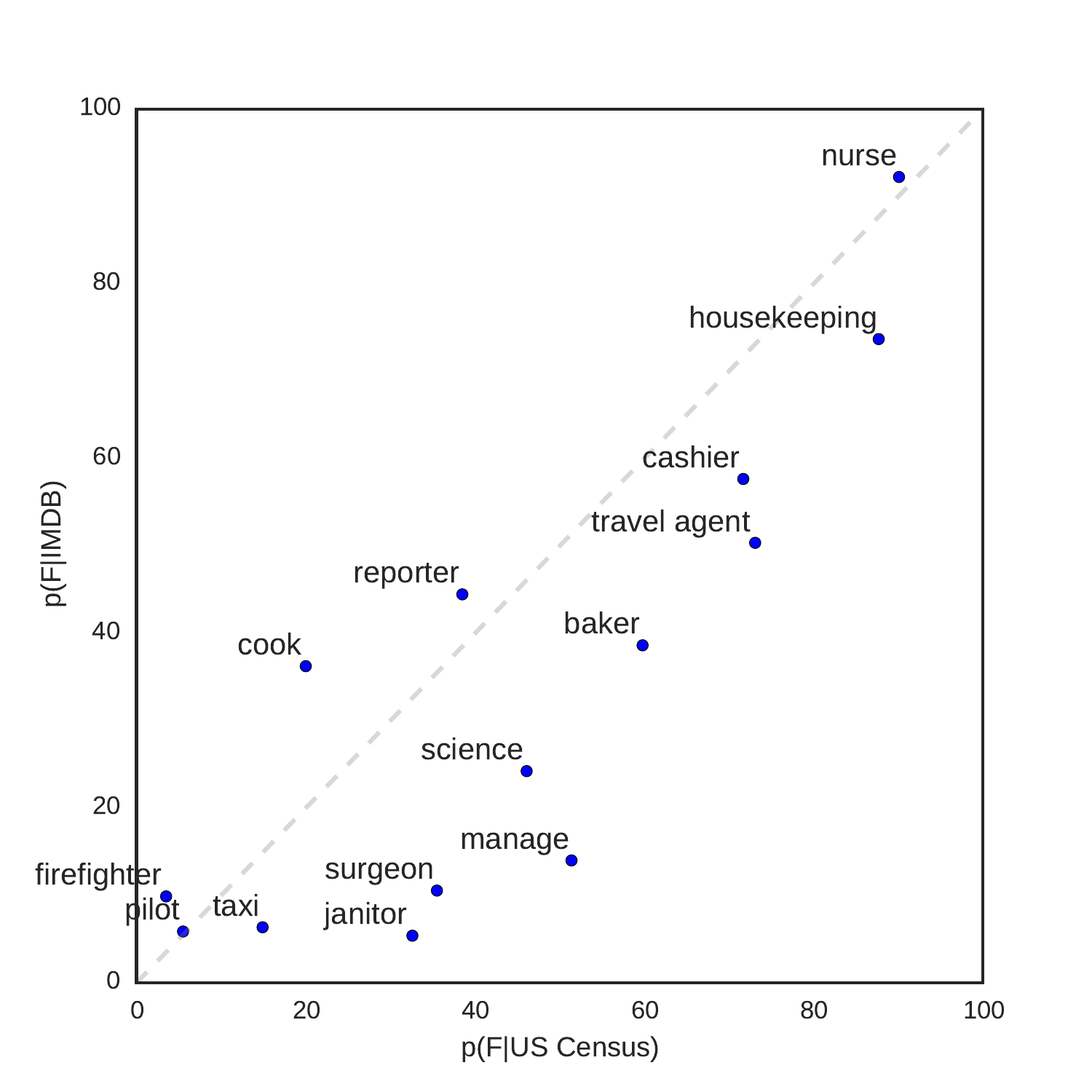}
  \caption{Proportion of female in movie dataset vs. US Census}
  \label{fig:movieVsReality}
\end{figure}

We can also examine role gender over time, searching for qualitative evidence that the gender associated with a specific role changes.
Figure \ref{gender-roles} shows the distribution of several roles, where we matched any role containing the query term (e.g. \txt{manage} would match \txt{bar manager}).
Onscreen \txt{nurses} have been traditionally almost uniformly female until the 1990s and now one in five nurses are played by male performers.
Conversely, the initial low proportion of onscreen female \txt{reporters} has risen and the proportion is now relatively even.

Our analyses to this point have only referenced \imdb data, but it is also interesting to examine how onscreen gender distributions compare with their real-world counterparts.
Figure \ref{fig:movieVsReality} shows how onscreen gender distributions map to those listed in the US Census\footnote{\url{http://www.bls.gov/cps/cpsaat11.pdf}}.
Points on the diagonal line have a portrayal consistent with the census distributions.
If a point is above the line (\txt{cook} and \txt{reporter}), then those roles are over-represented onscreen by female performers.
Conversely, points below the line suggest an under-representation onscreen by female performers.
For example, \txt{janitors}, \txt{surgeons} and \txt{managers} are mostly played by male performers in contrast to census data.
We also see under-representation of \txt{science}, \txt{baker} and \txt{cashiers}.
There are several limitations of this analysis.
Firstly, comparing user-generated roles with strict census roles introduces bias since we selected the mapping and selected roles.
Linking roles from the different sources to a common ontology would present a useful way to reduce manual effort in this step.
Secondly, we do not distinguish between US productions and those from other countries, so comparing with the US Census may introduce some noise.
Despite these factors, this analysis lets us draw an interesting counterpoint between onscreen gender representation and real-world figures.

\section{Conclusion}
Future work would concentrate on refining the data processing and adding useful structure for more rigorous statistical analysis.
This includes linguistic analysis to aggregate role synonyms, many of which are multi-word expressions.
Discriminating between media types (film, television) and genres may reveal interesting disparities on the gender proportion in them.
Identifying a production country would also be useful for analysis and language identification.
The \imdb data release does not report this information directly and it would have to be inferred.
Our current model emphasises the importance of secondary characters and treats them equally.
Extracting their roles from other data sources such as plot summaries or reviews would allow us to include major character roles and may motivate a ``central role'' weighting scheme.
Finally, contrasting on-screen gender representation with real data has the highest potential from a web science standpoint.
We provide exploratory analysis in Figure~\ref{fig:movieVsReality}, but further analysis must match the informal \imdb and formal census role ontologies.

%Most of the roles refer to secondary characters, as the primary ones will mostly be reported by the name of the character.
%However, we believe that this background cast actually provides a better insight into potential role biases, as the selection of who should portray the main roles is the product of a well-thought process where the gender of the actor plays an important role.
%Less time is spent in selecting secondary characters and implicit biases can therefore filter easier.
This paper presents methodologies for mining information about onscreen media gender from cast lists.
Despite the noise inherent in user-generated data, we assert that large-scale screen production metadata is a useful proxy for framing and answering questions about the evolution of roles over time, and how gender balances evolve.
We propose that the methodologies make for a compelling adjunct to traditional manual analyses and can help study how onscreen media is reflected onto the web, and eventually, how the web influences onscreen media.

\section{Acknowledgements}
The authors wish to thank \scterm{xrce} colleagues and Kellie Webster for thoughtful early feedback.

\clearpage
\bibliographystyle{abbrv}

%\bibliography{14-gender}  % sigproc.bib is the name of the Bibliography in this case
% ACM needs 'a single self-contained file'!
%
\end{document}